\documentclass[prd,aps,showpacs,tightline,twocolumn,nofootinbib]{revtex4}
\usepackage{graphicx}
\usepackage{amssymb}
\usepackage{amsmath}

\begin{document}

\title{Late-time Affleck-Dine baryogenesis after thermal inflation}
\author{Masahiro Kawasaki}
\author{Kazunori Nakayama}
\affiliation{Institute for Cosmic Ray Research,
The University of Tokyo,
Kashiwa 277-8582, Japan}
\date{\today}

\begin{abstract}
Thermal inflation can solve serious cosmological problems such
as  overproduction of gravitinos and moduli. 
However, it also dilutes the preexisting baryon asymmetry. 
We investigate a possibility that Affleck-Dine mechanism works after 
thermal inflation and generate the baryon number at an acceptable 
level using lattice calculation. 
We find that a proper amount of baryon number can be generated 
for appropriate model parameters. 
\end{abstract}

\pacs{98.80.Cq}
\maketitle

\section{introduction}

There are strong evidence for cosmological matter-antimatter asymmetry, 
or baryon asymmetry. Recent results from WMAP 3rd year data 
suggests~\cite{Spergel:2006}
\begin{equation}
	 \eta=\frac{n_B}{n_{\gamma}} \simeq (6.1 \pm 0.2) \times 10^{-10} .
\end{equation}
This value agrees well with the prediction of big-bang 
nucleosynthesis (BBN) inferred from 
observations of the light element abundances~\cite{Olive:2000}. 
Although various mechanisms are proposed in order to explain this 
value, it is yet unknown which is the right scenario. 
This is partly because more or less any baryogenesis mechanism 
is based on physics beyond the standard model 
and partly because thermal history of the universe before 
BBN is difficult to probe. 

Among many candidates, perhaps the best motivated model as physics 
beyond the standard model is supersymmetry (SUSY)~\cite{Nilles:1984}. 
Not only it naturally solves the hierarchy problem, but 
also provides the dark matter candidate. 
Furthermore, SUSY may play an important role in generating 
baryon asymmetry of the universe. In Affleck-Dine 
mechanism~\cite{Affleck:1985}
the dynamics of flat directions existing in the scalar potential 
of the minimal supersymmetric standard model (MSSM) can 
efficiently produce baryon asymmetry. 

On the other hand, the supersymmetry and 
supergravity theories predicts new particles 
such as gravitinos and moduli which interact with other particles 
only through gravity and hence have long lifetimes.
It is revealed that those particles may cause disaster for 
conventional cosmological scenario. 
Among such long-lived particles, the cosmological problem of the gravitino
has been most intensively studied by many authors.  
In gravity-mediation scenarios, gravitinos are expected to 
have masses of order 100 to 1000 GeV or so, and they decay
during or after BBN. 
Injected radiation and hadrons associated with the decay process 
significantly affect the light element abundances and spoil 
the success of BBN. 
In order to avoid this disaster, the abundance of
the garvitinos should be small enough.
Since the gravitino abundance is proportional to the reheating temperature
after inflation, the stringent constraint on the reheating temperature 
is imposed~\cite{Khlopov:1984,Kawasaki:1994af,Holtmann:1998gd,
Jedamzik:1999di,Kawasaki:2000qr,Kohri:2001jx,Cyburt:2002uv,Kawasaki:2004yh}.
In gauge-mediation scenarios, since mass of gravitino is estimated 
to be of order keV, the gravitino becomes LSP (lightest supersymmetric 
particle). In this case the condition that produced LSP should not 
overclose the universe also leads to constraint on the reheating
temperature~\cite{Moroi:1993}.  Furthermore, in the recent articles,
it was pointed out that the inflaton decays into gravitinos 
through supergravity effect, which results in overproduction 
of gravitinos even for low reheating temperature~\cite{Endo:2006qk}. 

More problematic situation arises when modulus fields exist. 
Potential of moduli is flat when SUSY is exact but SUSY breaking 
effects lift the flatness of the potential with curvature of order 
$m_{3/2}$.  In addition to this low energy supersymmetry breaking, 
vacuum energy from inflaton also breaks SUSY, which gives rise to 
modulus mass of the order of the Hubble parameter 
$H$~\cite{Copeland:1994}. 
For $H \gg m_{3/2}$  the modulus field sits down at the potential minimum
determined by Hubble induced mass term. 
This  minimum does not necessarily coincide with the 
true minimum of the potential, 
and naively it is expected that they are separated by amount 
of $M_P$($M_P$: Planck mass). 
When $H \sim m_{3/2}$, the modulus 
field starts to oscillate around the true minimum
with typical amplitude of $M_P$. 
Since the coherent oscillation of modulus fields behave as 
non-relativistic matter, it soon dominates the energy density 
of the universe. As the decay rate of moduli is of the same order 
as that of gravitinos, the same cosmological difficulty 
arises~\cite{Carlos:1993}. 
In contrast with the case of gravitinos 
lowering reheating temperature has nothing to do with 
moduli oscillation, so solving cosmological moduli problem 
is much more difficult.

So far several mechanisms to avoid the cosmological moduli problem 
have been proposed~\cite{Randall:1995,Linde:1996,Kofman:2004,MK&FT:2005,
Shuhmaher:2006,Yokoyama:2006,Lyth:1995}. Among them 
the most appealing solution relies on late-time entropy production 
to dilute the moduli. 
Here we concentrate on thermal inflation 
model~\cite{Lyth:1995,Yamamoto:1985rd} as 
a solution to the moduli problem.  In the thermal inflation scenario
a  scalar field called ``flaton''  causes a mini-inflation 
after the primordial inflation and the subsequent decay of the flaton 
produce huge entropy  enough to avoid the 
cosmological moduli problem (for detailed discussion, 
see~\cite{Asaka:1999}). It is noticed that the gravitinos are also
significantly  diluted by the thermal inflation and the gravitino 
problem is solved simultaneously.  

However, the same mechanism also dilutes the pre-existing 
baryon asymmetry. 
Even Affleck-Dine mechanism can not generate the enough 
baryon number which survives from thermal inflation~\cite{Kasuya:2002}. 
Thus, the baryon number should be regenerated after thermal inflation.
Since  thermal inflation ends at about $T\sim 1$~TeV, 
baryogenesis mechanism which works at sufficiently low energy 
is required. 
One possible candidate is electroweak baryogenesis~\cite{Trodden:1999}, 
but for the typical reheating temperature after flaton decay is 
lower than 100GeV, this mechanism may not work. 
Another possibility is modified version of Affleck-Dine 
baryogenesis with $LH_u$ flat direction which takes place after 
thermal inflation~\cite{Stewart:1996}. 
In Refs.~\cite{Stewart:1996,Jeong:2004} it was shown that the modified 
Affleck-Dine baryogenesis works by solving simplified dynamics of 
the scalar fields. 
However, actual dynamics of the scalar fields in this model is much
complicated and it is not clear whether the proper amount of baryon 
number is created or not when one solves full dynamics including 
all the relevant scalar fields. 
In this paper, therefore, we study the full dynamics of the scalar fields
using lattice simulation. We  adopt the model proposed 
in~\cite{Jeong:2004}.

This paper is organized as follows. 
In Sec.\ref{model} we describe our model and overview of the dynamics, 
in particular how the baryon asymmetry is generated in this model naturally. 
In Sec.\ref{num} we show the results of lattice simulation and 
constraints on the neutrino mass and $\mu$-term to obtain 
enough baryon number. In sec.\ref{conclusion} we give our conclusions.

\section{The model} \label{model}

\subsection{$\mu$-term and flaton field}

We start with the MSSM superpotential,
\begin{equation}
	W_{\rm MSSM}=y^u_{ij}Q_i H_u u_j + y^d_{ij}Q_i H_d d_j 
	+ y^e_{ij}L_i H_d e_j +\mu H_u H_d, \label{MSSMW}
\end{equation}
where $y$'s are Yukawa couplings, 
$i$ is the index of generation and $SU(2)$ or $SU(3)$ 
index is omitted. 
The last term of (\ref{MSSMW}) is called $\mu$-term where 
$\mu$ is the only dimensionful parameter in the MSSM. 
Hereafter, we also omit the generation indices for simplicity.
Furthermore, we add the following non-renormalizable superpotential 
which is responsible for neutrino mass:
\begin{equation}
	W_{\nu}=\frac{\lambda_{\nu}}{2M_\nu}(LH_u)(LH_u),  \label{numass}
\end{equation}
where $M_\nu$ denotes some cut-off scale. 
This term is regarded as Majorana mass term for left handed 
neutrino when $H_u$ has the VEV, 
and also be regarded as the term which lifts the potential 
of $LH_u$ flat direction~\cite{Gherghetta:1996}. 
We also introduce a singlet field $\phi$ which couples
with $H_u H_d$ as
\begin{equation}
	W_{\mu}=\frac{\lambda _{\mu}}{M^n}\phi ^{n+1} H_u H_d.
\end{equation}
Then this term explains
the natural value of $\mu$ with the expectation value of 
$\phi$~\cite{Kim:1984}, 
\begin{equation}
	\mu = \lambda_\mu \frac{\phi_0^{n+1}}{M^n}, \label{mu_phi0}
\end{equation}
where $\phi_0$ denotes the VEV of $\phi$. 
Since the required value of $\mu $ is around electroweak scale, 
this relation determines $\phi_0$. 
In our scenario the most attractive choice is $n=1$, because 
if we assume $M\sim M_P$ and $\lambda_{\mu}\sim O(1)$
eq.(\ref{mu_phi0}) would give 
$\phi_0 \sim 10^{10}$GeV, 
which is required for flaton field~\cite{Lyth:1995}. 
Therefore, in this model, the singlet field $\phi$, 
which is introduced to solve the naturalness of $\mu$-term, 
can be naturally regarded as flaton which leads to thermal inflation. 
In order to stabilize the flaton potential, we add the self coupling 
of the flaton field given by
\begin{equation}
    W_{\phi} = \frac{\lambda_\phi}{4M}\phi^4,
\end{equation}
where $\lambda_\phi \sim O(1)$.

To summarize, our superpotential is written as 
\begin{equation}
	\begin{split}
          W=& y^u QH_u u + y^d QH_d d + y^e LH_d e 
          + \frac{\lambda_\phi}{4M}\phi^4\\ 
         	    &+\frac{\lambda_{\nu}}{2M_\nu}(LH_u)(LH_u)  
                	+\frac{\lambda _{\mu}}{M}\phi ^{2} H_u H_d  
	      \label{fullW}.
    	 \end{split}
\end{equation}
We can forbid the other possible terms in the above superpotential 
using $R$-parity and some discrete symmetry, such as $Z_4$ symmetry, 
under which each field transforms non-trivially~\cite{Jeong:2004}. 
Furthermore, by gauging $Z_4$ symmetry we can avoid the domain wall 
problem associated with discrete minima of the potential. 
Here we only assume this problem can be avoided by some mechanism 
which does not affect the dynamics we are interested in.

Note that since the origin of neutrino mass term and flaton sector 
depend on different physics at high energy, 
cut-off scale $M$ and $M_\nu$ which appear in eq.(\ref{fullW}) 
do not need to coincide with each other.
In order to obtain a phenomenologically viable neutrino mass 
$m_\nu \lesssim 1$eV, The $M_{\nu}$ satisfies  
\begin{equation}
	M_\nu \gtrsim \lambda_\nu \times 10^{13}~{\rm GeV},
\end{equation}
which suggests that $M_{\nu} \ll M_P$ for the $O(1)$ coupling.
However, in this paper, we set $M=M_\nu \sim M_P$ and take 
$\lambda_\nu$ as a free parameter. 
Then  the natural parameter range for $\lambda_\nu$ is 
expected to be rather large compared with $\lambda_\mu$ 
or $\lambda_\phi $.

\subsection{Scalar field dynamics after thermal inflation}

During thermal inflation all fields are trapped to the origin 
due to thermal effects, and when thermal inflation ends at 
temperature $T \sim 100$~GeV, 
the flaton field $\phi$ and $LH_u$ flat direction begin to roll 
down to the instant minimum of their potential. 
The full scalar potential in our model consists of the $F$-term 
associated with (\ref{fullW}), 
$D$-term  and soft SUSY breaking terms. 
We can parameterize the flat directions $L H_u$ and $H_u H_d$ as 
\begin{equation}
    L = \begin{pmatrix}
         	0 \\ l
        \end{pmatrix},
    H_u=\begin{pmatrix}
            h_u \\ 0
        \end{pmatrix},
	H_d=\begin{pmatrix}
            0 \\ h_d
        \end{pmatrix}.
\end{equation}
The $F$-term potential is given by 
$V_F=\sum _\psi |\partial W/ \partial \psi|^2$, where $\psi$ runs 
all scalar fields included in the superpotential, explicitly, 
\begin{equation}
	\begin{split}
		V_F =& \frac{1}{M^2}\Bigl \{ | 
		       \lambda_\phi \phi^3+2\lambda_\mu \phi h_u h_d |^2
			+ | \lambda_\nu l h_u^2 |^2 \\
			&+ | \lambda_\mu \phi^2 h_d +\lambda_\nu l^2 h_u |^2  
		 	+ | \lambda_\mu \phi^2 h_u |^2  \Bigr \}.  \label{V_F}
	\end{split}
\end{equation}
The $D$-term contribution is written as 
\begin{equation}
	V_D=\frac{g^2}{2}
	   \bigl (  |h_u|^2 - | l |^2 - | h_d |^2 \bigr )^2.  
	   \label{V_D}
\end{equation}
Finally, the soft SUSY breaking terms including soft masses 
and A-terms are 
\begin{equation}
   \begin{split}
        	V_{\mathrm{SB}} =& V_0 - m_\phi^2 |\phi|^2 
	    + m_L^2| l |^2 - m_{H_u}^2| h_u |^2 
			   + m_{H_d}^2 | h_d |^2 \\
	   &+ \Bigl \{  \frac{A_\phi \lambda_\phi}{4M} \phi^4 +
			  \frac{A_\mu \lambda_\mu}{M}\phi^2 h_u h_d + 
			  \frac{A_\nu \lambda_\nu}{2M}l^2 h_u^2 
			  + \mathrm{c.c.} \Bigr \}  \label{V_SB}
   \end{split}
\end{equation}
Note that we assume that $m_{\phi}^{2} > 0$ and $m_L^2-m_{H_u}^2<0$ 
so that the flaton $\phi$ and the flat direction $LH_u$ rolls away from
the origin of the potential and create baryon number after thermal 
inflation. 
Justification of these ansatz is rather non-trivial~\cite{Casas:1996}, 
but it is possible to obtain valid parameter space~\cite{Jeong:2004}. 
In this model, as explained below, the most stringent constraint 
comes from the requirement that the true vacuum must break the 
electroweak symmetry spontaneously. 
This implies that mass matrix of the Higgs fields at the origin 
must have the negative determinant,
\begin{equation}
	(|\mu |^2-m_{H_u}^2)(|\mu |^2+m_{H_d}^2)<|B\mu |^2, 
	\label{constraint}
\end{equation}
where $B$ is given by 
\begin{equation}
	B = A_\mu + 2\lambda_\phi^* \frac{\phi_0^{*3}}{M\phi_0}
\end{equation}
from eqs.~(\ref{V_F}) and (\ref{V_SB}). 
Note that Hubble parameter $H$ is sufficiently 
smaller than the typical scale appearing in the dynamics and 
negligible in the interested regime.

Here we briefly describe outline of the dynamics of this model. 
As mentioned above, at the end of thermal inflation, 
the flaton $\phi$ and $LH_u$ 
flat direction begin to roll away from the origin. If the 
$LH_u$ direction first rolls away, the minimum of $LH_u$ is 
determined by the term 
$A_\nu \lambda_\nu l^2h_u^2/2M$ in eq.~(\ref{V_SB}). 
As $\phi$ rolls down the potential and increases its field value, 
$LH_u$ begins to feel the positive mass from the term 
$| \lambda_\mu \phi^2 h_u |^2 / M^2$ in eq.~(\ref{V_F}) and 
at the same time, the term 
$\lambda_\mu \lambda_\nu^* \phi^2h_dl^{*2}h_u^* / M^2 + h.c.$ gives 
$LH_u$ direction the angular kick. 
The minimum of the angular direction determined by this term 
is different from  the initial one, which is source of CP violation. 
If the following condition
\begin{equation}
	m_L^2 - m_{H_u}^2 + |\mu |^2 > 0  \label{positivemass}
\end{equation}
is satisfied, the potential for the $LH_u$ direction is stabilized 
at the origin and finally sit down there. 
However, it is not clear whether $U(1)$ conserving terms dominate 
the potential at this epoch. 
If not, $LH_u$ direction receives angular kick repeatedly and 
we lose the ability for predicting the resultant baryon number. 
Since the whole dynamics at this stage is quite complicated, 
we perform a numerical calculation based on lattice simulation.

\subsection{Reheating and baryon-to-entropy ratio}

Before going into the detailed analysis of the numerical calculation, 
we comment on the reheating and entropy production in our model. 
The final reheating temperature is determined by the flaton decay 
rate $\Gamma_\phi$ as
\begin{equation}
	T_R \sim \Bigl ( \frac{5}{4\pi^3 g_*} \Bigr )^{\frac{1}{4}}  
	\sqrt{\Gamma_\phi M_P}   
\end{equation}
where $g_*$ denotes the effective degree of freedom at $T_R$. 
Let us denote the epoch at which baryon asymmetry is generated 
as $t_B$ 
(which is almost the same time as the end of thermal inflation) 
and at which $\phi$ decays as $t_\phi$. 
The final baryon to entropy ratio is given by 
\begin{equation}
	\frac{n_B(t_\phi)}{s(t_\phi)} \sim 
	\frac{n_B(t_B)T_R}{m_\phi^2 \phi_0^2}. \label{baryon}
\end{equation}
Of course, it is lepton number that this process generates but
the electroweak sphaleron effect quickly converts the $B-L$ into $B$ 
number~\cite{Kuzmin:1985}.

Next we estimate the flaton decay rate. 
In the present model, $\phi$ can decay into two Higgs bosons 
or higgsinos if kinematically allowed. 
The decay rate is estimated as 
\begin{equation}
	\Gamma_\phi \sim 
	C \frac{|\lambda_\mu |^2 \phi_0^2 m_\phi}{M^2} 
	\label{decayrate}
\end{equation}
where $C$ is a constant of order $10^{-1} \sim 10^{-2}$. 
If we assume $M\sim M_P$, this gives $T_R\sim 1$GeV. 
But note that even if the above decay modes are kinematically 
forbidden, the flaton can decay into two photons or gluons 
through the one loop diagrams associated with the particles 
in the thermal bath required for thermal inflation~\cite{Asaka:1999}. 
These processes also give decay rate that is roughly the same order  
as eq.~(\ref{decayrate}). 
It should be noted that the natural range of the final reheating 
temperature is much lower than the electroweak scale.

\section{Numerical analysis of the dynamics} \label{num}

We have investigated the full dynamics with potential~(\ref{V_F}), 
(\ref{V_D}) and (\ref{V_SB}) using lattice simulation. 
In Refs.~\cite{Stewart:1996} and \cite{Jeong:2004},
an ad hoc large damping term  was put by hand under 
assumption that the flaton field quickly decays 
through parametric resonance.
In this work, we have not made any such  artificial assumption. 
The only non-trivial input for computing the dynamics is 
initial condition. As the initial condition, we set the initial
values of all fields 
at around 1TeV as homogeneous part, 
as is expected for thermal fluctuations at the end of thermal 
inflation. 
Our simulation is performed with one-dimensional lattice 
with 128 grid points.
Initial fluctuations come from quantum fluctuation around the homogeneous mode.
We apply the method used in LATTICEEASY \cite{Felder:2000}
for including these quantum fluctuation.
In order to eliminate unphysical effect due to large quantum 
fluctuations at short distance, we have cut the initial quantum 
fluctuations with mode $k>m $,
although this does not much affect the result
(for more detail about initial condition for tachyonic potential, 
see~\cite{Felder:2001} ).

It is found that the dynamics is rather sensitive to the model 
parameters. However, 
since there are many parameters in the model and full 
parameter search is beyond the scope of this paper,
we have performed the simulations fixing the most of parameters as
$m_\phi =180$~GeV, $m_{H_u} =700$~GeV, 
$m_{H_d} =800$~GeV, $m_L =640$~GeV, 
$\lambda_\phi=4$, $A_\mu =450$~GeV, $A_\nu=200$~GeV, $A_\phi =20$~GeV, 
and arg$(\lambda_\phi \lambda_\mu^*)=-\pi/4$.
We also take $M=M_P$. 
All $A$-terms are taken to be real by field redefinition and 
hence the only remaining parameter associated with $CP$ angle is 
arg$(\lambda_\mu \lambda_\nu^*)$.
The other parameters are varied in each of the following analysis.

\begin{figure}[htbp]
	\begin{center}
		\includegraphics[width=1.0\linewidth]{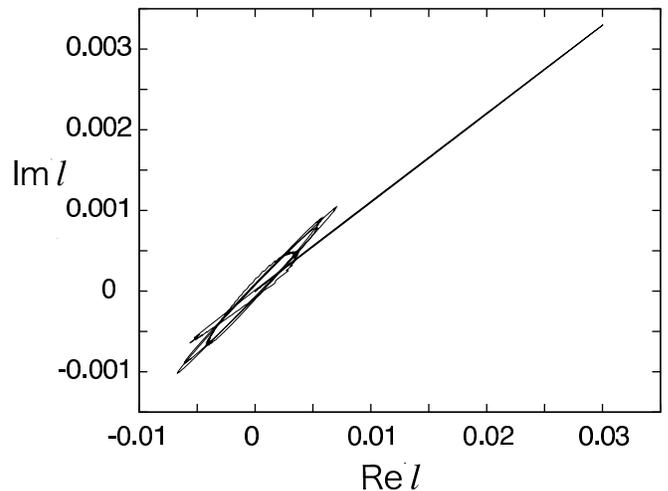}
		\caption{Typical motion of the field $l$ in the complex plane. 
		         Here we write only zero mode. 
		         The field value is normalized by $10^{9}$GeV.
		         We take $|\lambda_\mu |=35, |\lambda_\nu |=10^4$, 
		         arg$(\lambda_\mu \lambda_\nu^*)=\pi/16$. }
		\label{fig:Rel_Iml}
	\end{center}
\end{figure}

First, we show the typical motion of the field $l$ 
in Fig.~\ref{fig:Rel_Iml} when only the homogeneous mode is 
taken into account.
We can see that first $l$ rolls down to the displaced minimum and 
then pull back to the origin, as explained in the previous section. 
In this process, $l$ is kicked to angular direction, 
and finally the field rotates around the origin with constant 
angular momentum (i.e., conserved lepton number), 
in the complex plane. 


\begin{figure}[htbp]
	\begin{center}
		\includegraphics[width=1.0\linewidth]{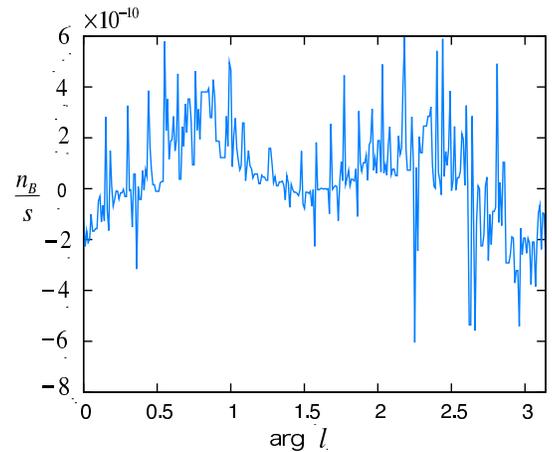}
		\caption{Initial angular dependence of baryon asymmetry $n_B/s$,
				when we take $T_R=1$GeV. 
				The same parameters as in Fig.\ref{fig:Rel_Iml} are taken. }
		\label{fig:argl_nL}
	\end{center}
\end{figure}

\begin{figure}[htbp]
	\begin{center}
		\includegraphics[width=1.0\linewidth]{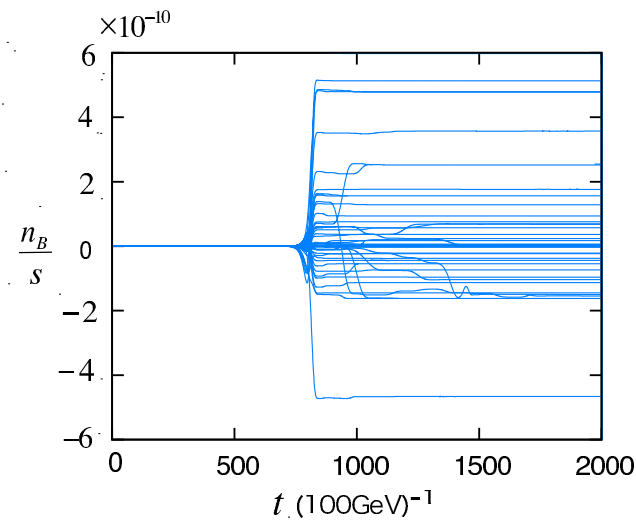}
		\caption{Time dependence of baryon number. 
				$t$ is normalized by (100GeV)$^{-1}$.
		        		Parameters are same as Fig.\ref{fig:argl_nL}, 
				and each line corresponds to different initial angle. 
				Data points are reduced in order to make it easy to see.}
		\label{fig:t_nL}
	\end{center}
\end{figure}

From Fig.~\ref{fig:Rel_Iml} it is obvious that lepton number is 
really generated. However, the present universe (or at BBN or 
recombination epoch) contains 
large number of causally disconnected regions at the era of 
thermal inflation. 
This means that the real baryon number is average over many regions 
with different initial values of $l$ field. 
The initial values of $l$ is determined by the thermal 
fluctuations at the end of thermal inflation. Thus, the initial values 
of phase of $l$ is random and $|l|$ also fluctuates around 
$\langle |l| \rangle \sim T$. Taking fluctuations of $l$ into account,
we perform the simulations varying the initial phase of $l$
with initial $| l |$ fixed. 
As for the initial value of $\phi$, we fix its phase,
but we have confirmed that initial angular 
dependence of the $\phi$ field does not much affect the subsequent 
dynamics.
We show in Fig.~\ref{fig:argl_nL} the relation between the initial 
angle arg($l$) and the resultant baryon asymmetry. 
From this figure it is seen  that the average baryon number is really 
non-zero. 
Fig.~\ref{fig:t_nL} shows the time evolution of the baryon number. 
From this figure we can clearly see the baryon number is 
finally conserved. 


\begin{figure}[tbp]
	\begin{center}
		\includegraphics[width=1.0\linewidth]{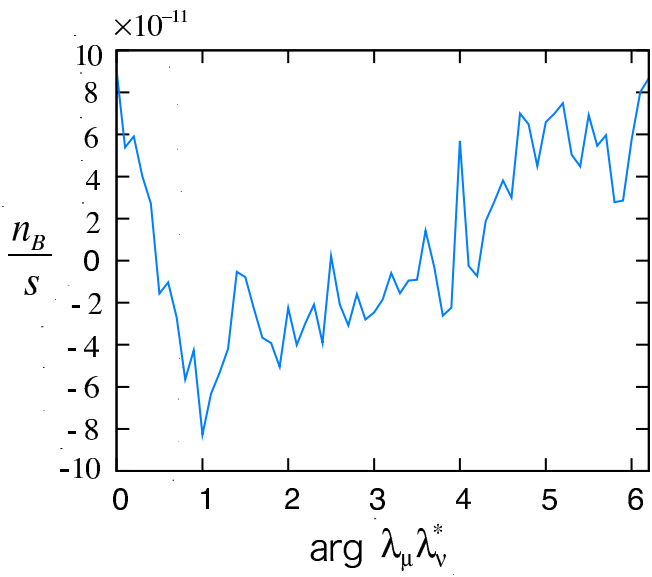}
		\caption{Relation between arg$(\lambda_\mu \lambda_\nu^*)$ 
		and net baryon number. 
		When $CP$ is conserved, the net lepton number becomes zero 
		as explained in the text.}
		\label{fig:CP_nL}
	\end{center}
\end{figure}

As explained in previous section, whether or not the net 
baryon asymmetry is generated depends on CP phase. 
The phase of the flaton field takes the value that minimizes 
the potential of the angular direction determined by the term 
$A_\phi \lambda_\phi \phi^4/M$ in the interested time scale.
The relevant potential of angular direction for $LH_u$ 
comes from the term $A_\nu \lambda_\nu l^2 h_u^2$ initially, 
and subsequent dynamics depends on the terms
$\lambda_\mu \lambda_\nu^* \phi^2 h_d l^{*2}h_u^*$ 
and  $\lambda_\phi \lambda_\mu^* \phi^3 \phi^* h_u^* h_d^*$. 
Initially, the angular minimum lies in the direction 
arg$(lh_u)=\pi/2$ and $3\pi/2$. 
The angular minimum is unchanged when 
arg$(\lambda_\mu \lambda_\nu^*)=\pi/4$ and $5\pi/4$ 
so the net baryon number expected to become zero.
This is seen in Fig.~\ref{fig:CP_nL}.

\begin{figure}[bp]
	\begin{center}
		\includegraphics[width=1.0\linewidth]{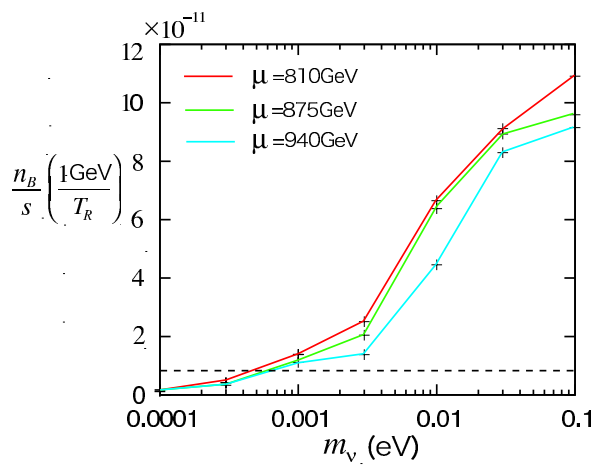}
		\caption{Relation among baryon number $(n_B/s)\times (T_R/\mathrm{GeV})^{-1}$, 
		$m_\nu$(eV) and $\mu$(GeV). 
		Region below the horizontal line requires $T_R \gtrsim 10$GeV to 
		obtain desired value of $n_B/s$. }
		\label{fig:mnu_mu_nL}
	\end{center}
\end{figure}

Finally, we fix arg$(\lambda_\mu \lambda_\nu^*)=\pi/16$ and 
choose freely  the absolute value of $\lambda_\mu$ and 
$\lambda_\nu$. Each of these parameters is directly related 
to $m_\nu$ [eq.~(\ref{numass})]and $\mu$ [eq.~(\ref{mu_phi0})]. 
In Fig.\ref{fig:mnu_mu_nL} we plot generated baryon number 
$(n_B/s)\times (T_R/\mathrm{GeV})^{-1}$
in terms of $m_\nu$ and $\mu$. 
Since our choice of arg$(\lambda_\mu \lambda_\nu^*)=\pi/16$ 
corresponds to nearly maximum CP phase,
the baryon-to-entropy ratio in Fig.\ref{fig:mnu_mu_nL}
is reduced for different choice of 
arg$(\lambda_\mu \lambda_\nu^*)$. Thus, the apparent large $n_B/s$
is not a problem.   
For smaller $m_\nu$ (or $\lambda_\nu$ ) than $10^{-3}$~eV, 
$T_R$ must exceed about 10GeV even for maximum CP phase,
which is invalid for the present model.
For smaller value of $\mu$ ( or $\lambda_\mu$) than about $800$~GeV, 
although the constraint~(\ref{positivemass}) can be satisfied,
we could not get appropriate dynamics  
due to smallness of positive mass squared of $LH_u$ direction. 
On the other hand, the constraint~(\ref{constraint}) 
invalidates $\mu$ larger than about 
840GeV. 
As a result, only constrained parameter region 
around $\mu \sim 800 - 840$GeV and 
$10^{-3}$eV$ \lesssim  m_\nu \lesssim 10^{-1}$eV
can survive.

However, we have to mention that our model has quite large 
number of parameters and the dynamics is sensitive to them. 
Thus the stringent constraints should not be taken too seriously.
Complete analysis of parameter dependence is beyond the scope of 
the present paper,
but it should be noticed that our results indicate that 
it is really possible to generate baryon asymmetry even 
after thermal inflation.

\section{Conclusions} \label{conclusion}

In this paper detailed analysis of the Affleck-Dine mechanism after 
thermal inflation has been performed. 
The present model includes only one additional gauge singlet which does 
not exist in the MSSM.  This singlet field explains naturally both 
the origin of $\mu$-term and thermal inflation.
This is an appealing feature when considering the cosmological 
moduli problem seriously.
However, it is not trivial matter to generate enough baryon 
asymmetry consistent with late-time entropy production. 
In this paper we have demonstrated that
a proper amount of baryon 
asymmetry can be generated using the lattice calculation 
for the whole dynamics without any artificial assumption.
Although the parameter tuning is necessary for the present
mechanism to work, 
It is noticed that since the dynamics is highly sensitive to parameters 
in the model there can be another parameter sets appropriate 
for our purpose which we could not find.

We stress that if the late time entropy production takes place and 
dilute the preexisting baryon number density, 
the modified Affleck-Dine process considered in this paper
is the only known  mechanism to re-create the baryon number 
at very low energy scale after late-time entropy production.
It is interesting that this mechanism can work with quite low reheating temperature
even below the electroweak scale. 
This opens up a new possibility for such a low-energy 
baryogenesis, and may be tested in future accelerator experiments.

\end{document}